# Role of Cold Storage in Rooting of Stem Cuttings: A Review

## Aram Akram Mohammed [a*]

[a] *Department of Horticulture, College of Agricultural Engineering Sciences, University of Sulaimani, Sulaimanyah, Kurdistan Region, Iraq.*

## ABSTRACT

Cold storage is a strategy mainly used for ornamental cuttings for the purpose of prolonging production duration that provide the cuttings with the best condition at the time of planting. The results have been obtained from cold storage of carnations and chrysanthemums confirmed the fact that the herbaceous cuttings could be stored for about two months at not lower than 0ºC and more than 4ºC, with similar results of the cuttings directly planted without storage. On the other hand, hardwood cuttings are stored at low temperature in order to improve rooting ability via reducing rooting inhibitors that occur in dormant cuttings which were taken during dormant seasons (late fall or early winter). So, declining in rooting inhibitors has been observed in the hardwood cuttings of deciduous trees and *Vitis spp* which were subjected to low temperature for around 2-6 months. Consequently, rooting was improved in these species. Also, cold storage of cuttings of some coniferous trees increased rooting at a high rate. Other factors like cultivar and storage method may interact with cold storage duration and temperature degree as well. This review article outlines the information of the studies that have been obtained as a result of cold storage of the cuttings of different species.

## 1. INTRODUCTION

Vegetative propagation by stem cuttings decisively depends on the ability of the stem cuttings to produce adventitious root (AR) because cuttings contain vegetative buds which emerge easily as shoots, but root formation is challenging. Adventitious root formation (ARF) is controlled by many endogenous and exogenous components. Having a thorough knowledge of these components will be helpful for manipulating rooting in cuttings. Some of the factors that induce rooting cuttings are specific to one step of ARF rather than all steps, however, others influence rooting from the beginning to the end [1]. Manipulation of the exogenous environment of the cuttings is sometimes conducted in order for certain purposes. This will have a direct reflection on the capacity of the cuttings to form roots. In this context, cuttings of different genera are stored at low temperature for a variety of reasons [2]. Mainly, herbaceous cuttings of some ornamentals are held at low temperature during excessive production which would permit propagators to balance market and schedule production [3]. On the other hand, cuttings of trees and shrubs expose to cold storage aims to improve rooting, due to the changes occurred inside the cuttings during cold storage [4,5]. Cuttings of different species differently responded to cold storage depending on the type of cutting, storage duration, and the temperature degree of storage. In this review effects of cold storage on cuttings before rooting are discussed which were obtained in different studies.

## 2. IMPACT OF COLD STORAGE ON HERBACEOUS CUTTINGS OF CARNATION AND CHRYSANTHEMUM

Storing rooted and unrooted cuttings of ornamental plants for long periods allows the growers to increase production capacity since propagation could be done when space, labor, and market demand are in proper condition [6]. Besides, Vošnjak and Osterc [7] indicated that cold storage of the cuttings gives opportunities to transport the cuttings a long way and extend their shelf life via slowing down metabolism, which preserves growth potential and cuttings quality. Variable results have been obtained when cuttings of ornamental species are stored at low temperature. The factors which have a role in maintaining cutting shelf life are temperature degree, storage period, cultivar, light, etc. In their pioneer work, Holley and Farmer [8] found that carnation cuttings could store at -1 to 4.4 °C for 73 to 113 days. The stored cuttings gave rooting similar to fresh cuttings (73 to 93%). They further referred that the cuttings should be in a sealed covering, not moisturized and dipped in water at the time of storage. In the following studies on cold storage of carnation cuttings, it was revealed that the effect of cold storage duration (2-12 weeks) at 4±2°C were the same on the final rooting percentage of carnation cuttings taken from three cultivars ('Oriana', 'Elsy' and 'Virginie'), whether they were treated or untreated with auxins [3]. On the contrary, rooted and non-rooted cuttings of two carnation cultivars (Dianora and Vittorio) were assessed to determine cold storage duration (0 - 4 months) at 0 - 0.5°C. It was obtained that Vittorio was better than Dianora for resisting cold storage if it was rooted or non-rooted, and the rooting rate reduced from 100% in the cuttings without storage (control cuttings) to 58% in the cuttings stored for 4 months [9]. Moreover, the results of cold storage of four chrysanthemum cultivars ('Emily', 'Naomi', 'Anna', and 'Debonair') by Rajapakse et al. [10] precisely showed that temperatures below zero (−1 and −3°C) caused freezing damage to the cuttings. While, at 0 or 3°C and dark storage 'Anna', and 'Debonair' gave better results than of 'Emily', 'Naomi', when they were stored for 4 to 6 weeks, including maintaining quality, shortening days from planting to anthesis, retaining a high level of soluble sugars. They further found that cold storage in light caused leaf necrosis to decline in the stored cuttings of the cultivars. In another study was conducted on the chrysanthemum cuttings from two cultivars ('Puma' and 'Cassa'), it was obtained that from leaves of the stored cuttings at 0.5 - 5°C for 2, 3, and 4 weeks, starch entirely disappeared within 2 weeks, and sugar concentrations were decreased as well, but did not impede ARF [11]. Furthermore, two groups of chrysanthemum cuttings were received irradiation with 0 to 70 Gy of gamma-ray before and after cold storage at 4°C for 4 weeks. The group that was exposed to irradiation in advance of storage was better at rooting [12]. De Almeida et al. [13] stored chrysanthemum cuttings of Super White, Sheena, Dark Orange Reagan, and Town Talk cultivars at 7°C for 0, 1, 2, 3, 4, 5, and 6 weeks in winter and summer. The data proved that the cuttings taken in winter could be stored for a short period (2 weeks), in contrast, the ones from summer were stored for up to 4 weeks without notable shortcomings. The best results of storing cuttings of chrysanthemum from the Tzevlev cultivar were achieved when they were stored for one month rather than 2, 3, and 4 months at 4°C [14].

## 3. RESPONSE OF DECIDUOUS HARDWOOD CUTTINGS TO COLD STORAGE

The buds of deciduous trees enter dormancy in winter. This is concurrent with variation in phytochemicals and phytohormones inside the tree, which have an apparent reflection on the ability of the cuttings of these trees to form adventitious roots. During the dormant state, growth inhibitors increase (abscisic acid) and growth promoters decrease (auxin, cytokinin, and gibberellin), at the earlier stages [15]. The growth inhibitors are not favourable at the time of rooting the cuttings. So, it has been detected that subjecting the tree and the cuttings of deciduous species to low temperature reduces the growth inhibitors and elevates the growth promoters [16]. Cold storage of the cuttings is a treatment used to facilitate rooting in hardwood cuttings of deciduous trees. In this context, the inhibitors of indole acetic acid (IAA) action were found in the extract of fresh cuttings of chestnut. These inhibitors were nullified after exposing the cuttings to cold storage at -10°C for 5 months. At the same time, vanillyl and salicyl alcohols which are rooting inducers increased as well [17]. At breaking dormancy state in late winter, it was seen that sycamore cuttings root so readily. Therefore, sycamore hardwood cuttings were taken in late fall and treated with indole butyric acid (IBA), sucrose, and captan then stored for 1 month at 4 °C. The results explained that the cuttings were treated with IBA, sucrose, and captan before storing gave the best rooting than those treated without cold storage or treated after cold storage [18]. Additionally, hardwood cuttings of poplar were stored in two different methods, heeled-in outdoor on November 14 and refrigerated at -2 and 2°C (to stay dormant, the heeled-in ones were lifted and stored at 2°C until the time of planting). On May 23, the cuttings were planted in a nursery. The cuttings stored at -2°C showed the best rooting (82%), and 73% for those stored at 2°C, but heeled-in resulted in the lowest rooting (60%) [19]. In the continuation of the studies were conducted on cold storage of hardwood cuttings of chestnut, it was reported that 4 months of storage at 4°C from December to April was favourable to form root in the cuttings treated with 4000 mg.L$^-$ IBA before storage. Also, the extract of the cold-stored cuttings induced rooting in *Phaseolus* cuttings bioassay, which was evidence that rooting inhibitors tested before storage were replaced by rooting promotors during cold storage [20]. Despite the beneficial effects of cold storage on hardwood cuttings, there were some reports about the ineffectiveness or negative consequences of cold storage on hardwood cuttings of some species. Regarding, Tsipouridis et al. [21] found in hardwood cuttings of peach that were stored for 2 months in a refrigerator at 2 - 4°C that the cuttings differently responded to cold storage according to tree age from which the cuttings were taken. The cuttings taken from 1-year-old trees positively reacted to the cold storage, whereas the rooting rate was decreased in the stored cuttings taken from 11-year-old trees. Moreover, Hawramee [22] identified in the cuttings of black locust that storing the cuttings for 3 and 6 weeks at 5°C adversely affected the rooting of the cuttings more than those without cold storage.

## 4. ROOTING BEHAVIOUR OF HARDWOOD CUTTINGS OF *Vitis spp.* AFTER COLD STORAGE

In the studies that have been carried out to determine role of cold storage in the rooting of hardwood cuttings of *Vitis spp*, it was resolved that cold storage has affirmative consequences on the rooting and some hormonal and phytochemical changes in the cuttings. One study showed that the condition and period of storing of *Vitis* cuttings had an evident role in the rooting of the cuttings. The cuttings were collected in January and stored outdoor in sand, then planted on April 14 rooted better than the ones refrigerated at 0 - 2.2°C and planted on February 15, March 15, or April 14 [23]. According to Blennerhassett and Considine [24], the rooting of *V. champinii* cv. Ramsey was more challenging early in the dormancy period, whereas became less challenging with prolonged subject to freezing temperatures. In view of the hormonal changes that occurred in the cuttings at the time of cold storage, Skene [25] noted that the cytokinin ratio detected in xylem sap of the vine canes stored at 1°C for 6 months was high compared to the canes from actively growing plants without cold storage. The sap of the stored canes induced more callus formation in soybean which was used as a bioassay. Additionally, soluble carbohydrates and rooting capacity in the cuttings of *Vitis* 140 Ruggeri rootstock were investigated by comparing the cuttings collected in October and stored at 2°C with those taken monthly directly from the field (from December to April) without storage. The obtained results explained that cold storage mobilized soluble carbohydrates and caused earlier rooting as well [26]. In a greenhouse experiment in which rooting of *Vitis aestivalis* Michx. Norton cuttings were evaluated under intermittent mist, more roots were formed in the cuttings stored at 5°C for 55 days after collecting in January [27]. Furthermore, cold storage grants the growers a longer time for prolonging the propagation period of *Vitis spp* cuttings. Sabir and Sabir [28] summarized that they could store cuttings of seven grapevine cultivars of *Vitis vinifera* at 1 °C for 8 months in a clean, cool room. However, the viability of the cuttings was adversely influenced when they were stored in a damp sand medium for a long term. Hence, the response of the cultivar was variable to the cold storage method and duration. In spite of all these, Buck [29] indicated in his study that hardwood muscadine cuttings (*Vitis rotundifolia* Michx.) which received 4°C for 30 days in advance of planting gave rise to diminish rooting rate from 27% to 19%.

## 5. COLD STORAGE OF SOME CONIFERS AND OLIVE CUTTINGS

For the purpose of raising rooting, the cuttings of some conifers were stored at low temperature in some different studies. In this respect, cuttings of *Pinus radiata* D. Don took benefits of cold storage when they were stored for 20 to 50 days to accelerate rooting and increase root number per cutting [30]. Likewise, for the cuttings of *Pinus sylvestris*, which were taken from a 5-year-old dormant, winter-chilled tree and then exposed to short-day treatment accompanied by receiving 0±5°C for 60 days, the superior rooting was achieved [31]. The relationship between cold storage and water potential in the cuttings of *Pinus taeda* was tested [32]. The results illustrated that cold storage for short time (at 30°C for 7 hours or at 4 °C for 21 hours) was satisfactory to preserve the severed cuttings from moderate levels of water stress. In another study, the correlation between chlorophyll fluorescence (Fv/Fm) and rooting in cuttings of 10 *Taxus* cultivars was examined. For this reason, the cuttings were stored at 2.5 or 5°C in the dark, after 4 weeks of storage the cuttings were treated with auxins (1700 mg·L$^{-1}$ IBA and 1100 mg·L$^{-1}$ NAA). During cold storage, the Fv/Fm was reduced, but the normal range was reached after planting in the propagation bed for several weeks. However, chlorophyll fluorescence and rooting were not strongly correlated [33]. Enhancing rooting in the cuttings of *Pinus halepensis* was focused on by storing them for 4 weeks at 4°C, before planting. The data on rooting percentage clarified that rooting heightened as a result of cold storage, three times more than of the cuttings without cold storage [34]. Apart from the conifers, the semi-hardwood cuttings of olives were taken in October, treated with 2500 ppm IBA, and then stored for 3 days at 4°C. The greatest rooting percentage was achieved as a result of the storage [35]. Besides, El-Sherif [36] investigated the impact of cold storage at 4°C for 3 and 6 days on rooting of leafy olive cuttings, from different cultivars. He also found that the cuttings were treated with 3500 ppm IBA and then stored for 3 days showing the best rooting percentage, total indoles, and soluble sugar contents. Storage for 6 days decreased the rooting rate.

## 6. CONCLUSION

The results of the researches reviewed in this study showed that ornamental herbaceous cuttings could be stored for a long time (4 - 10 weeks), generally at 0 - 4°C without adverse effects. However, slight different consequences were noted related to cultivar. Also, temperature below zero was not convenient to store ornamental herbaceous cuttings. Besides, cold storage effectively improved rooting in the dormant hardwood cuttings of deciduous trees, which were collected in late fall or early winter and stored at about -2 to 4°C for about 2 - 5 months, depending on tree species. Whereas, temperature above 4°C and storage period of more than 5 months along with storage methods negatively affect the cuttings. Refrigeration had more advisable than holing or heeled-in damp sand outdoor. The consequences of cold storage of hardwood cuttings of *Vitis spp* showed that the best cold storage was between 0 to 2°C for about 6 months. The latest studies didn't recommend storing the dormant cuttings of *Vitis spp* cuttings in damp sand outdoor as well. On the other hand, according to the investigation conducted on evergreen coniferous cuttings, cold storage of the *Pinus spp* cuttings taken in winter and stored for around 40 - 50 days at 4°C favourably enhanced rooting. In contrast, cuttings of broad-leaf evergreens such as olive could not be stored for more than 3 days. Crucially, if rooting promotors are used for the stored cuttings, should be used in advance of the storage rather than after storage.

## COMPETING INTERESTS

Author has declared that no competing interests exist.